\documentclass[graybox]{svmult}

\usepackage{amsmath}
\usepackage{amssymb}

\usepackage{mathptmx}       
\usepackage{helvet}         
\usepackage{courier}        
\usepackage{type1cm}        
\usepackage{makeidx}         
\usepackage{graphicx}        
\usepackage[bottom]{footmisc}

\makeindex             

\begin{document}

\title*{The environment of barred galaxies revisited}
\author{Bernardo Cervantes Sodi, Cheng Li, Changbom Park  \& Lixin Wang}
\institute{Bernardo Cervantes Sodi \at Korea Institute for Advance Study, Dongdaemun-gu, Seoul 130-722, Republic of Korea, and Centro de Radioastronom\'ia y Astrof\'isica - UNAM, Apdo. Postal 3-72 (Xangari), 58089 Morelia, Mich., Mexico, \email{b.cervantes@crya.unam.mx}
\and Cheng Li \& Lixin Wang \at Key Laboratory for Research in Galaxies and Cosmology of Chinese Academy of Sciences, Shanghai Astronomical Observatory, Nandan Road 80, Shanghai 200030, China
\and Changbom Park \at Korea Institute for Advanced Study, Dongdaemun-gu, Seoul 130-722, Korea }
%
%
\maketitle

\abstract*{We present a study of the environment of barred galaxies using galaxies drawn from the SDSS. We use several different statistics to quantify the environment: the projected two-point cross-correlation function, the background-subtracted number count of neighbor galaxies, the overdensity of the local environment, the membership of our galaxies to galaxy groups to segregate central and satellite systems, and for central galaxies we estimate the stellar to halo mass ratio (M$_{\mathrm{*}}/$M$_{\mathrm{h}}$). When we split our sample into early- and late-type galaxies, we see a weak but significant trend for early-type galaxies with a bar to be more strongly clustered on scales from a few 100 kpc to 1 Mpc when compared to unbarred early-type galaxies. This indicates that the presence of a bar in early-type galaxies depends on the location within their host dark matter halos. This is confirmed by the group catalog in the sense that for early-types the fraction of central galaxies is smaller if they have a bar. For late-type galaxies, we find fewer neighbors within $\sim50$ kpc around the barred galaxies when compared to unbarred galaxies from the control sample, suggesting that tidal forces from close companions suppress the formation/growth of bars. For central late-type galaxies, bars are more common on galaxies with high M$_{\mathrm{*}}/$M$_{\mathrm{h}}$ values, as expected from early theoretical works which showed that systems with massive dark matter halos are more stable against bar instabilities. Finally, we find no obvious correlation between overdensity and the bars in our sample, showing that galactic bars are not obviously linked to the large-scale structure of the universe.}

\abstract{We present a study of the environment of barred galaxies using galaxies drawn from the SDSS. We use several different statistics to quantify the environment: the projected two-point cross-correlation function, the background-subtracted number count of neighbor galaxies, the overdensity of the local environment, the membership of our galaxies to galaxy groups to segregate central and satellite systems, and for central galaxies we estimate the stellar to halo mass ratio (M$_{\mathrm{*}}/$M$_{\mathrm{h}}$) . When we split our sample into early- and late-type galaxies, we see a weak but significant trend for early-type galaxies with a bar to be more strongly clustered on scales from a few 100 kpc to 1 Mpc when compared to unbarred early-type galaxies. This indicates that the presence of a bar in early-type galaxies depends on the location within their host dark matter halos. This is confirmed by the group catalog in the sense that for early-types, the fraction of central galaxies is smaller if they have a bar. For late-type galaxies, we find fewer neighbors within $\sim50$ kpc around the barred galaxies when compared to unbarred galaxies from the control sample, suggesting that tidal forces from close companions suppress the formation/growth of bars. For central late-type galaxies, bars are more common on galaxies with high M$_{\mathrm{*}}/$M$_{\mathrm{h}}$ values, as expected from early theoretical works which showed that systems with massive dark matter halos are more stable against bar instabilities. Finally, we find no obvious correlation between overdensity and the bars in our sample, showing that galactic bars are not obviously linked to the large-scale structure of the universe.}

\section{Background}
\label{sec:1}

\begin{figure*}[t]
  \begin{center}
\includegraphics[width=\hsize]{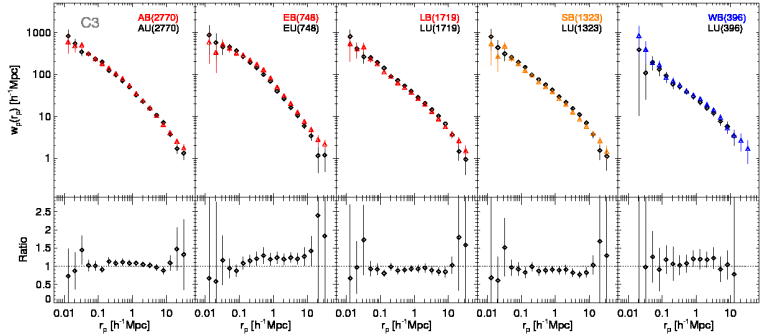} 
 \end{center}
    \caption{Projected  2PCCF for  barred  $w_p^{bar} (r_p)$  and
      unbarred  $w_p^{unbar}  (r_p)$  galaxies (top  panels)  and
      $w_p^{bar} (r_p)$ to $w_p^{unbar} (r_p)$ ratio (bottom
      panels) for the different subsamples of our control sample C3 where
      the galaxies share a  common distribution of stellar mass $M_*$,
      color g$-$r, and stellar surface mass density $\mu_*$.
      Panels  from left  to right correspond  to the  whole galaxy
      sample of  barred (AB orange triangles) plus  unbarred (AU black
      diamonds) galaxies, early-types barred (EB orange triangles) and
      unbarred  (EU  black  diamonds),  late-types barred  (LB  orange
      triangles)  and unbarred  (LU black  diamonds),  strongly barred
      late-types  (SB  yellow triangles)  and  LU,  and weakly  barred
      late-types (WB blue triangles) and LU.   }
\label{fig:2PCCF_C3}
\end{figure*}

With the outcome of large galaxy surveys, the role of environment in triggering or suppressing the formation of stellar bars in galaxies has been a popular topic of research, with a large variety of results, depending on the methods employed \cite{vandenBergh02, Li09, Lee12, Skibba12}. 

In this work we employ a large volume-limited sample of $\sim$30,000 galaxies to study the likelihood for galaxies to host bars as a function of environment, using several statistics to study different scales, with the aim of giving a complete study for galaxies in the nearby universe. For a more detailed description of the sample and a complete discussion of the results, we refer the reader to \cite{Lin14, Cervantes-Sodi14}.

\section{Main Results}
\label{sec:2}
\begin{figure*}[t]
  \begin{center}
\includegraphics[width=\hsize]{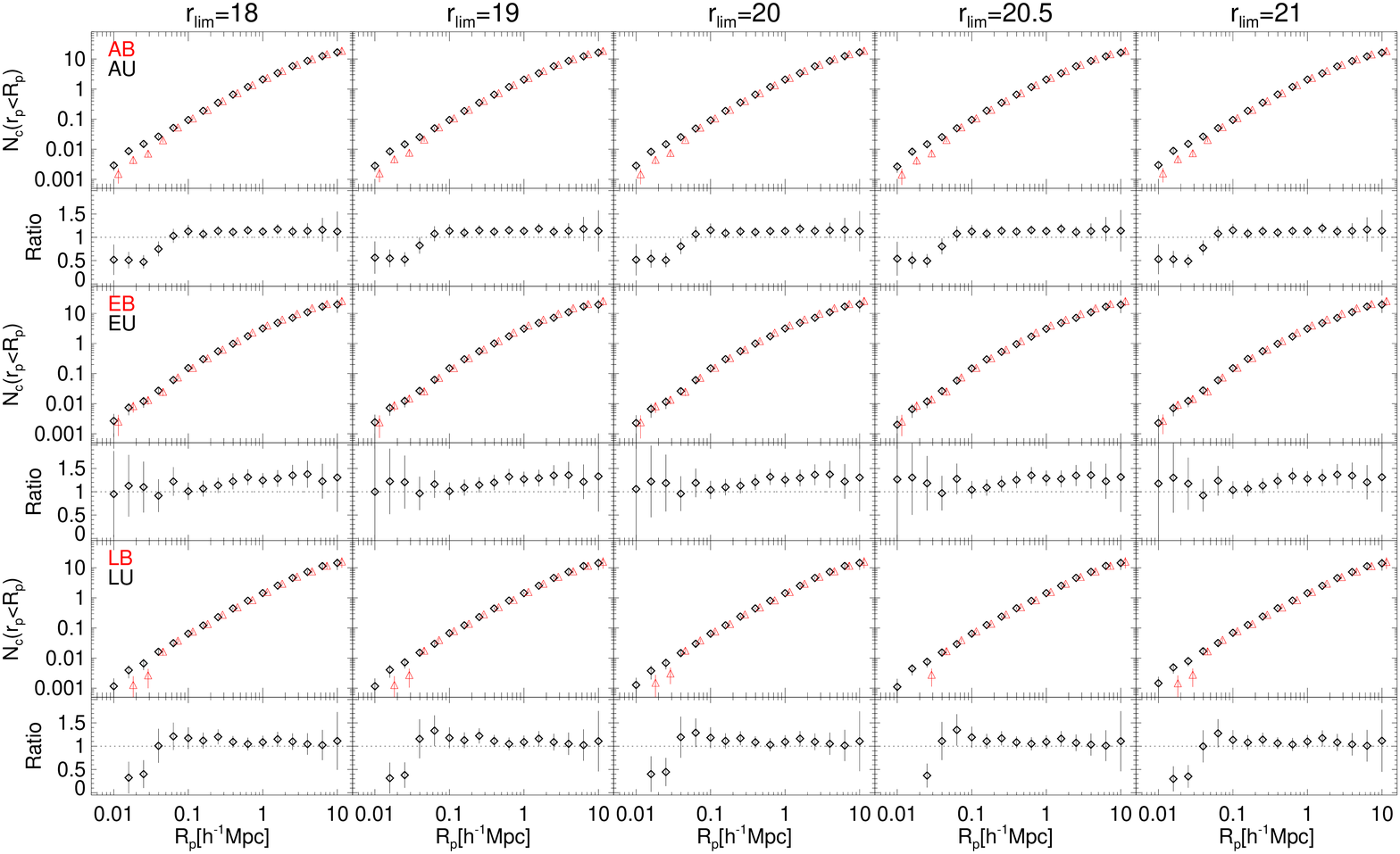} 
  \end{center}
    \caption{Average  counts  of galaxies  in  the photometric  sample
      within a given  projected radius R$_P$ from the  galaxies in
     our samples of barred and unbarred control samples .
      Each line  corresponds  to different  apparent
      magnitude  limits in r-band  ($r\leq$ 18,19,20,20.5,21)  for the
      galaxies in the photometric sample.}
  \label{fig:nc}
\end{figure*}

Our first approach is to compute the two-point cross correlation function of our galaxy sample with respect to a reference sample of the general galaxy population (2PCCF,  $w_p(r_p)$),  from scales of a few tens of kiloparsecs up to a few tens of megaparsecs. We then construct a control sample of unbarred galaxies with matched stellar mass, color and surface mass density ( $\triangle$log$M_* \leq$ 0.08 , $\triangle(g-r)\leq$ 0.025,
$\triangle$log$\mu_*\leq$ 0.08). In Figure 1, first panel, we present the 2PCCF for the barred sample (AB), the control sample (AU) as well as the ratio between them. In this case, no significant overclustering is detected for the full sample of barred galaxies over unbarred ones, but if we segregate the samples by morphology, the 2PCCF for early-type barred galaxies (EB) show a weak but significant trend of overclustering on scales from a few 100 kpc to 1 Mpc, when compared with unbarred early-type galaxies (EU), a trend not found with the late-type galaxies, regardless if their bars are classified as strong or weak. This result implies that the presence of bars in early-type galaxies depends on the location of the galaxies within their host dark matter halos.

With the fiber collision problem suffered by the SDSS at small separation
angles, we explore the clustering of our samples at small separations by
computing  the background-subtracted  neighbor  counts, $N_C(r_p<R_p)$,
which  is the  number  of galaxies  in the  photometric
reference sample  within the projected  radius $R_p$ of the  barred or
unbarred  galaxies,  with  the   effect  of  chance  projection  being
statistically corrected \cite{Lin14}. In Figure~\ref{fig:nc}  we plot $N_C$,
panels top to bottom show the results for the whole
sample, the early-type and late-type subsamples respectively
and  from left to right,  the  results  for different  apparent  $r$-band  limiting
magnitude applied  to the  photometric reference sample,  ranging from
$r_{lim}=18$  for  the  left-most   panels  to  $r_{lim}=21$  for  the
right-most panels. For early-type galaxies, the overclustering of barred
galaxies becomes again evident for scales larges than 100 kpc, but
this time the subsample including only late-type galaxies shows a lack
of neighbors in the vicinity of barred galaxies on scales 
$\lesssim$ 50  kpc when compared with unbarred galaxies,
suggesting that tidal forces from close companions suppress
the formation/growth of bars.

\begin{figure*}
\begin{tabular}{ccc}
\includegraphics[width=.33\textwidth]{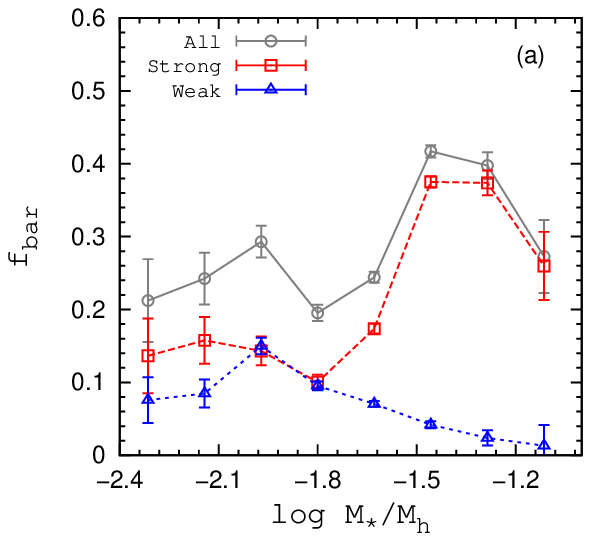} &
\includegraphics[width=.33\textwidth]{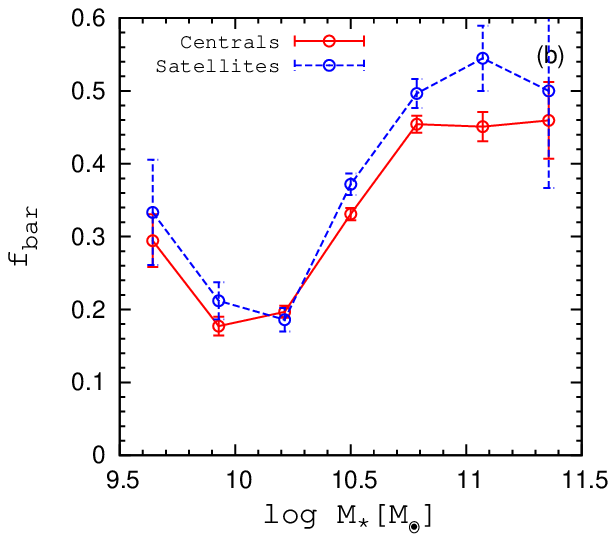} &
\includegraphics[width=.33\textwidth]{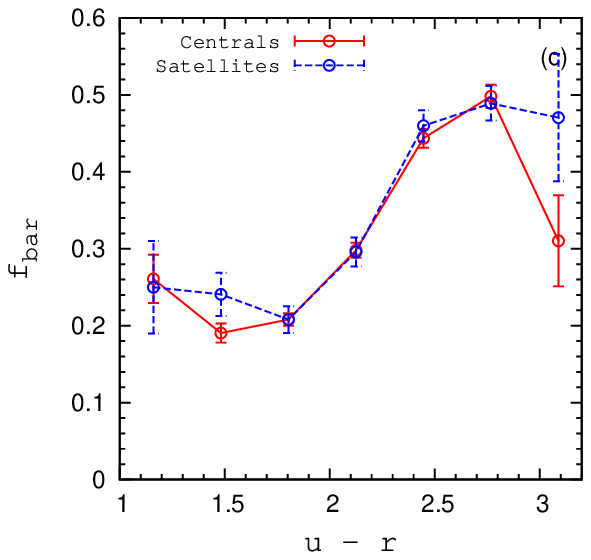} \\

\end{tabular}
\caption{The fraction of barred galaxies $f_{\mathrm{bar}}$ as a function of
\textit{(a)} stellar-to-halo mass ratio, for strong, weak and strong plus weak bars, \textit{(b)} stellar mass for central and satellite galaxies and, \textit{(c)} color for centrals and satellites.}
  \label{fig3}

\end{figure*}

Using the galaxy catalog by \cite{Yang07}, we classified our galaxies
as central or satellite galaxies. For the late-type centrals, we estimate the
stellar-to-halo mass ratio (M$_{\mathrm{*}}/$M$_{\mathrm{h}}$),
using the stellar mass data from the
MPA/JHU SDSS database \cite{Kauffmann03} and the halo masses from the same galaxy catalog.
Figure~\ref{fig3} (a) shows that the bar fraction is a strong function of this ratio, with strong bars
more commonly found in galaxies with high M$_{\mathrm{*}}/$M$_{\mathrm{h}}$,
as expected form theoretical works which show that galaxies with massive dark
matter halos are more stable against bar formation, while weak bars follow the opposite trend.
 
With our galaxies segregated into centrals and satellites, we compare the bar fraction
of both subsamples as a function of stellar mass and color in Figure~\ref{fig3} (b,c). At fixed stellar mass
 we find that the bar fraction of satellite galaxies is slightly higher than that of
central galaxies, but at fixed color this difference vanishes. We interpret this as
follows; the color of satellite galaxies, on average, is redder than the color of
centrals at the stellar mass ranges involved in our study. Given that the bar
fraction is higher for redder galaxies, the difference of the bar fraction between
centrals and satellites is not directly due to the group environment, but indirectly
through its dependence on internal morphology.

\end{document}